\documentclass{article}
\usepackage{amsmath,amsfonts,amsthm}
\usepackage{empheq,cases}
\usepackage{xcolor}
\usepackage{graphicx}
\usepackage{subfig}
\usepackage{algorithm}
\usepackage{algpseudocode}
\usepackage{url}
\usepackage[hidelinks]{hyperref}      

\def\Col{{\mathrm{Col}}}

\def\cI{{\cal I}}
\def\cQ{{\cal Q}}
\def\bR{{\mathbf{R}}}
\def\bE{{\mathbf{E}}}
\def\Prob{{\mathrm{Prob}}}
\def\Argmin{\mathop{\mathrm{Argmin}}}
\def\Argmax{\mathop{\mathrm{Argmax}}}

\def\argmax{\mathop{\mathrm{argmax}}}
\def\Diag{{\mathrm{Diag}}}
\theoremstyle{remark}

\newtheorem{proposition}{Proposition}

\DeclareMathOperator{\Tr}{Tr}

\newcommand{\lam}{\lambda}

\newcommand{\beq}{\begin{equation}}
\newcommand{\eeq}{\end{equation}}
\newcommand{\be}{\begin{enumerate}}
\newcommand{\ee}{\end{enumerate}}
\newcommand{\bi}{\begin{itemize}}
\newcommand{\ei}{\end{itemize}}

\newcommand{\mycomment}[1]{}

\title{Radiation design in computed tomography via convex optimization }
\author{
Anatoli Juditsky
\thanks{LJK, Universit\'e Grenoble Alpes, 700 Avenue Centrale,  38401 Domaine Universitaire de Saint-Martin-d'Hères, France
{\tt anatoli.juditsky@univ-grenoble-alpes.fr}}
\and Arkadi Nemirovski
\thanks{Georgia Institute
 of Technology, Atlanta, Georgia
30332, USA, {\tt nemirovs@isye.gatech.edu}}
\and
Michael Zibulevsky\thanks{Department of Computer Science,
Technion---Israel Institute of Technology,
Haifa 32000, Israel, {\tt mzib@cs.technion.ac.il} \newline
This work was supported by Multidisciplinary Institute in Artificial intelligence MIAI {@} Grenoble Alpes (ANR-19-P3IA-0003) and by the Israel Council For Higher Education---Planning \& Budgeting Committee}}
\begin{document}
\maketitle

\begin{abstract}
\noindent Proper X-ray radiation design (via dynamic fluence field modulation, FFM) allows to reduce effective radiation dose  in computed tomography without compromising image quality. It takes into account patient anatomy, radiation sensitivity of different organs and tissues, and location of regions of interest. We account all these factors within a general convex optimization framework.
\end{abstract}

\section{Introduction}
{Recently,}
there has been significant research interest in dynamic fluence
field modulation (FFM), {which consists in varying} the beam
shape throughout the scan {allowing for the} adaptation of the spatial x-ray
distribution to conform to the patient anatomy {(cf., e.g., \cite{gang2017task} and references therein)}. A range of
hardware solutions have been proposed, including dynamic
wedges \cite{szczykutowicz2013design,hsieh2013feasibility,liu2013dynamic}, 
 fluid-filled chambers  \cite{peppler1982digitally, szczykutowicz2015fluid, shunhavanich2015fluid, liu2014dynamic}, 
and slitbased multiple aperture devices \cite{stayman2016fluence, mathews2017experimental}. 

Until now, {to the best of} our knowledge, the problem of FFM radiation planning has been treated within non-convex optimization framework, and only rough distribution of radiation with relatively small number of parameters  (coefficients of basis functions)  was modeled.
In this work we utilize convex optimization techniques to optimize the fine structure of the fluence, namely, the amount of radiation sent into each projection bin (towards the corresponding detector pixel).\footnote{Following the publication of our paper on arXiv, it came to our attention that a concept akin to ours was previously outlined in an unpublished Master's thesis ``Optimal Anisotropic Irradiation For X-Ray Computed Tomography'' by Y. Chernyak, Technion, October 2021; see the Technion Graduate School website: 
\href{https://graduate.technion.ac.il/en/department-en-4msc/}{https://graduate.technion.ac.il/en/department-en-4msc/}}
\par
The main body of this paper is organized as follows. In Section \ref{sect2} we present the CT observation model underlying our {derivations},
introduce and motivate the {\sl loss index} of radiation design responsible for the asymptotical, as the total amount of radiation grows,
quality of the Maximum Likelihood recovery of the image of interest. We then pose the problem of optimal radiation design as the problem of optimizing the loss index under general convex constraints on the design. The resulting optimization problem is convex and thus efficiently solvable (at least in theory, and to some extent, also in practice). In Section \ref{sect3}, we report on the (in our appreciation, encouraging)
results of a ``proof of concept'' numerical experiment implementing the radiation design proposed in Section \ref{sect2}. Section \ref{sect4} contains some concluding remarks.

\section{CT {Radiation} Design}\label{sect2}
\def\Poisson{{\mathrm{Poisson}}}
\subsection{CT observation scheme}
The CT observation scheme we intend to consider is as follows:
\begin{enumerate}
\item $x\in\bR^n_+$ is the ``signal'' -- the discretized body attenuation density, so that $n$ is the number of voxels in the {field} of view;
\item $w\in\bR^m$ is the vector of observations; observations are indexed by bins, $m$ being the total number of bins;
\item $A\in\bR^{m\times n}$ is the ``projection matrix'';
\item $q\in\bR^n_+$ is the  vector with entries $q_i$ which are the expected numbers of photons sent to bin $i$ during the stude.
\par
We assume that the $i$-th observation stemming from signal $x$ is
\begin{equation}\label{aeq1}
\omega_i\sim \Poisson(q_i\exp\{-[Ax]_i\},
\end{equation}
where $\Poisson(\mu)$ is the Poisson distribution with parameter $\mu\geq0$:
$$\Prob_{\iota\sim\Poisson(\mu)}\{\iota= k\}={\mu^k\over k!}e^{\mu},\; k=1,2,...
$$
We assume also that entries $\omega_i$ in the vector of observations $\omega$ are independent across $i=1,...,m$.
\item {Given observation $\omega$ stemming from the unknown signal $x$, we want} to recover the vector $Bx$, where $B\in\bR^{\nu\times n}$ is a given matrix.\footnote{In our experiments  $Bx$ is a truncation of  image $x$ towards the region of interest (ROI).}
{We} measure the recovery error in the standard Euclidean norm $\|\cdot\|_2$ on $\bR^\nu$.
\end{enumerate}

\subsection{Radiation sensitivity - from voxels to bins }
Informally, our goal is to find radiation design $q$, which provides best image quality given the whole-body effective dose. Various organs and tissues have different sensitivity to radiation---effective dose caused by unitary radiation exposure. This can be reflected in 3D voxel-wise sensitivity map $s$, which is standardly  used in radiation therapy planning.

{We assume that,} in addition, an (approximate) attenuation map of the body {is available}, {and} we can compute an auxiliary  bin-sensitivity vector $c$ with entries $c_i$ reflecting effective dose obtained from a unit of radiation sent into bin $i$. Constant whole body effective dose  is provided by the condition
\beq \label{cq}
c^T q = \mbox{const},
\eeq
Thus our goal is  to find radiation design $q$, which provides the best image quality under constraint \eqref{cq}.

\subsection{Maximum Likelihood estimate} With $A$ and $q$ given, we intend to recover $Bx$ as $B\widehat{x}(\omega)$, where $\widehat{x}(\omega)$ is the Maximum Likelihood (ML) estimate. Denoting by $a_i^T$, $i=1,2,...,m$, the rows of $A$, the log-likelihood of observing $\omega$, the signal being $x$, is
$$
\begin{array}{rcl}
{\cal L}_q(\omega,x)&=&\sum_{i=1}^m[\omega_i\ln(q_i\exp\{-a_i^Tx\})-q_i\exp\{-a_i^Tx\}-\ln(\omega_i!)]\\
&=&\sum_{i=1}^m[-\omega_ia_i^Tx -q_i\exp\{-a_i^Tx\}]+R(q,\omega).\\
\end{array}
$$
Consequently, maximizing ${\cal L}_q(\omega,x)$ in $x$ is the same as maximizing in $x$ the concave in $x$ function
\begin{equation}\label{aeq2}
L_q(x)-[\sum_i\omega_ia_i]^Tx,\;\; L_q(x)=-\sum_iq_i\exp\{-a_i^Tx\}
\end{equation}
and is therefore an efficiently solvable problem; the resulting estimate
$$
\widehat{x}(\omega)\in\Argmax_x\left[L_q(x)-[\sum_i\omega_ia_i]^Tx\right]
$$
of $x$  is a function of $\omega$ depending on $q$ as a parameter.
\subsection{Radiation Design via Loss index {minimization}}
Our goal is ``presumably good,'' in terms of the performance of the resulting estimate, selection of $q$ in a given compact convex set ${\cal Q}\subset\bR^m_+$. This informal goal can be {stated formally} in different ways; the formalization we propose is {based on the following observation}.
\par
A. In the ``noiseless case'' where the observations are $\omega_i=\mu_i$, $\mu_i=q_i\exp\{-a_i^Tx_*\}$, $x_*$ being the true signal (instead of being Poisson random variables with parameters $\mu_i$), the ML estimate exactly recovers the true signal.  In other words, setting  $\gamma_*=\sum_i\mu_ia_i$,  we get $x_+\in\Argmax_x[L_q(x)-\gamma_*^Tx]$. Indeed, since $L_q$ is concave and smooth in $x$, it suffices to verify that $\nabla_xL_q(x_*)=\gamma_*$, which is evident.
\par
B. Now assume that the positive semidefinite {Hessian} matrix
$$
H:=-\nabla^2L_q(x_*)=\sum_i\mu_ia_ia_i^T
$$
is positive definite, or, which is the same when $q>0$, that $A$ is of rank $m$. Let $\overline{L}_q(\cdot)$ be the second order Taylor approximation of $L_q(\cdot)$ around $x=x_*$. Were we specifying our estimate of $x$ as
\[\widetilde{x}=\widetilde{x}(\omega):=\argmax_x[\overline{L}_q(x)-[\sum_i\omega_ia_i]^Tx],\]
we would have
$$
\nabla_x\overline{L}_q(\widetilde{x})=\sum_i\omega_ia_i,\,\nabla_x\overline{L}(x_*)=\sum_i\mu_ia_i,
$$
that is, $\widetilde{x}-x_*=-H^{-1}\underbrace{\sum_i[\omega_i-\mu_i]a_i}_{\delta}$. Hence, taking into account the immediate relation $\bE\{\delta\delta^T\}=H$ (recall that $\omega_i\sim\Poisson(\mu_i)$ are independent across $i$),
$$
\begin{array}{rcl}
\bE\{\|B\widetilde{x}-Bx_*\|_2^2\}&=&\bE\left\{\delta^TH^{-1}B^TBH^{-1}\delta\right\}
=\bE\left\{\Tr\left(H^{-1}B^TBH^{-1}\delta\delta^T\right)\right\}\\
&=&\Tr\left(H^{-1}B^TBH^{-1}\bE\left\{\delta\delta^T\right\}\right)=\Tr\left(H^{-1}B^TB\right)\\
&=&\Tr(BH^{-1}B^T).
\\
\end{array}
$$
This computation suggests, as a  meaningful  ``loss index'' of $q$ (the smaller it is, the better for our purposes) the function
\begin{equation}\label{aeq3}
{\cal I}_*(q)=\Tr\left(B\left[\sum_iq_i\exp\{-a_i^Tx_*\}a_ia_i^T\right]^{-1}B^T\right)
\end{equation}
Needless to recall, the above reasoning is informal---our estimate of $x$ is not $\widetilde{x}$, it is $\widehat{x}$. {Nevertheless, it} can be proved that when $A$ is of rank $m$ and $q$ becomes large, specifically, $q=t\bar{q}$ with $\bar{q}>0$, when $t\to\infty$, our informal reasoning yields correct asymptotics of the expected squared $\|\cdot\|_2$-error of the recovery of $Bx$\footnote{Namely, with $q=t\bar{q}$, $\bar{q}>0$, one has  $\bE\{\|B\widehat{x}-Bx_*\|_2^2\}=(1+o(t))t^{-1}{\cal I}_*(\bar{q})$ as $t\to\infty$.}. We, however, neither present these asymptotic results, nor need them---our goal at this point is  to find a convenient criterion to be optimized over $q\in{\cal Q}$ in order to get a presumably good measurement design, and to this end no rigorous justification of our choice is needed. What indeed is important for us, is that ${\cal I}_*$ is a convex function of $q>0$, and thus is well suited for minimization.
\par
A bad news is that strictly speaking, we cannot use ${\cal I}_*$ as a criterion to be minimized---this quantity depends, as on a parameter, on the unknown to us true signal $x_*$. {On the other hand,} when replacing in the right hand side of (\ref{aeq3}) the unknown quantities $\rho^*_i:=\exp\{-a_i^Tx_*\}$ with their approximations $\rho_i$ which are  within factor $c$ from $\rho_i^*$, that is $c^{-1}\rho^*_i\leq\rho_i\leq c\rho_i^*$,
the resulting right hand side in (\ref{aeq3}) is within the same factor from the true one. Therefore, assuming that an approximate  attenuation map $\rho$ is available a priori, or that it can be estimated using observation $\overline{\omega}$ from a pilot run with, say, equal to each other ``moderate''  $\overline{q}_i=\overline{q}$, we may  select $q$ by minimizing over ${\cal Q}$ the observable criterion
\begin{equation}\label{aeq4}
{\cal I}(q)=\Tr\left(B\left[\sum_iq_i\rho_ia_ia_i^T\right]^{-1}B^T\right).
\end{equation}
This is, essentially, the course of actions we propose.
\subsection{Regularization}\label{alternate}
The problem of minimizing $\cI(q)$ over $q\in\cQ$ is convex and as such can, theoretically,  be solved to whatever high accuracy in a computationally efficient manner. At the same time, numerical experimentation shows that this problem
may happen to be ill-conditioned, resulting in difficulties when solving it numerically. To overcome, to some extent, these difficulties, we can replace minimizing over $q\in\cQ$ the objective $\cI(q)$ given by (\ref{aeq4}) with minimizing over the same domain the penalized objective
\begin{equation}\label{aeq4pen}
\cI_\lambda(q)=\Tr\left(B\left[\sum_iq_i\rho_ia_ia_i^T+\lambda I_n\right]^{-1}B^T\right)
\end{equation}
where  regularizing coefficient $\lambda$ is nonnegative; a good value of this coefficient could be found experimentally. Thus, in the sequel we intend to optimize measurement design by solving convex optimization problem
\begin{equation}\label{optprob}
\min\limits_{q\in\cQ}\left\{\cI_\lambda(q):=\Tr\left(B\left[\sum_iq_i\rho_ia_ia_i^T+\lambda I_n\right]^{-1}B^T\right)\right\}
\end{equation}
which can be solved by standard Convex Optimization algorithms. It turns out, however, that one can utilize problem's structure to design a dedicated algorithm which, as our experiments show, is much faster than the ``general purpose'' algorithms. The {following} underlying observation is essentially well known:
\begin{proposition}\label{propo} Let $P$ be an $m\times n$ matrix of rank $n$, $B$ be a $k\times n$ matrix, and $q,\rho$ be positive $m$-dimensional vectors. Then, setting $W=W[q]:=\Diag\{q_i\rho_i,i\leq m\}$,
\[\hspace{-4cm}
\cI_\lambda(q):=\Tr\left(B\left[\sum_iq_i\rho_ia_ia_i^T+\lambda I_n\right]^{-1}B^T\right)
\]
\begin{subequations}
\begin{numcases}{=}
\min_{G\in\bR^{k\times m}}\left[\lambda^{-1}\|GA-B\|_F^2+\Tr(GW^{-1}G^T)\right],&\text{$\lambda>0$} \label{poslambda.a}\\
\min_{G\in\bR^{k\times m}}\left[\Tr(GW^{-1}G^T):GA=B\right],&\text{$\lambda=0$}\label{poslambda.b}
\end{numcases}
\end{subequations}
where $\|\cdot\|_F$ is the Frobenius norm
\end{proposition}
We provide the proof of Proposition \ref{propo} in Appendix \ref{app.prop1} for the sake of completeness.
\paragraph{Alternating minimization.}
Proposition \ref{propo} suggests the {\sl alternating minimization} algorithm for minimizing $\cI_\lambda(q)$ over $q\in\cQ$. Assume for the sake of definiteness that $\lambda>0$ (modification for the case of $\lambda=0$ is immediate).
By (\ref{poslambda.a}), the problem of interest (\ref{optprob}) can be rewritten equivalently as
$$
\min_{G\in\bR^{k\times n},q\in\cQ} \Phi(G,q):=\left[\lambda^{-1}\|GA-B\|_F^2 +{\Tr(GW^{-1}(q)G^T)}\right]
$$
To solve the latter problem, we start with some positive $q^0\in\cQ$. At a step $t$ of the algorithm, given positive $q^{t-1}\in\cQ$,
\begin{itemize}
\item we minimize $\Phi(G,q^{t-1})$ over $G\in\bR^{k\times m}$; this is {an unconstrained minimization problem with strongly convex quadratic objective,} and its solution $G_t$ is given by explicit formula:
\beq
G_t= B[A^TW[q^{t-1}]A+\lambda I_n]^{-1}A^TW[q^{t-1}];
\label{G_update}
\eeq
\item after $G_t$ is computed, we specify $q^t$ by minimizing $\Phi(G_t,q)$ over $q\in\cQ$, which boils down to finding
$$
q^t\in\Argmin_{q\in\cQ}\underbrace{{\sum}_{j=1}^m{\|\Col_j(G_t)\|_2^2\over q_j\rho_j}}_{=\Tr(G_tW^{-1}[q]G_t^T)}.
$$
When $\cQ$ is simple, the latter problem is simple as well. For instance, in the case of $\cQ=\{q\in\bR^m_+:c^Tq\leq1\}$ with $c>0$, assuming that $G_t$ has no zero columns (which can be ensured by an arbitrarily small perturbation of $G_t)$, the solution is given by
\beq
[q^t]_j={r_j/c_j\over c^Tr},\,r_j=\|\Col_j[G_t]\|_2\sqrt{c_j/\rho_j},\,1\leq j\leq m.
\label{q_update}
\eeq
\end{itemize}
After $G_t,q_t$ are computed, we pass to step $t+1$.

\section{Computational experiment}\label{sect3}
\begin{figure}[ht!]
\centering
\includegraphics[width=0.99\columnwidth]{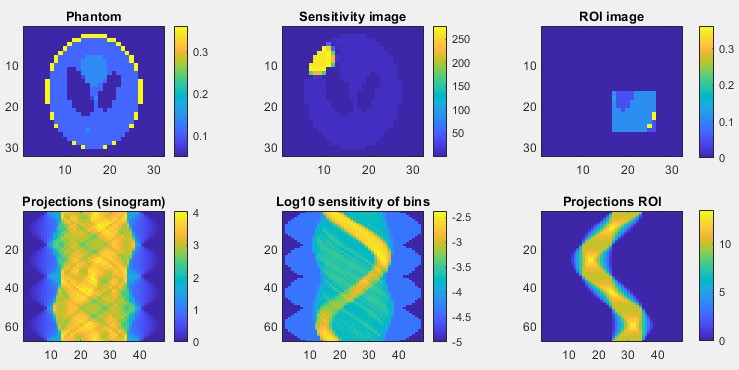}
\caption{Top row: phantom, sensitivity image and region of interest (ROI). Bottom row: phantom sinogram, sensitivity of the bins and sinogram of the ROI. \label{fig_phantom}}
\end{figure}
Here we report on a toy ``proof of concept''  numerical experiment on recovering $n=32\times 32$-pixel Shepp-Logan type phantom with 68 projection angles spread uniformly over 360 degrees.
\subsection{Experimental setup}
Unlike conventional tomography, where 180 degree angle span is sufficient, we consider 360 degree span to account for the fact that effective radiation dose is generally different when photons are sent along the same line from opposite directions due to different attenuation between the source and a particular body point.
In our experiments, the number of bins is $m=2828$;
we utilize the baseline (uniform) radiation design $\overline{q}_i\equiv 2$e5 photons/bin.
 Figure \ref{fig_phantom}, top row, shows the  attenuation map, radiation sensitivity, and region of interest (ROI) which we aim to reconstruct. Bottom row shows the corresponding maps in the sinogram (projection) space.\footnote{When computing sinograms we used a partially modified ``Matlab demo for 2-D tomographic reconstruction'' toolbox by L. Han, see \href{https://github.com/phymhan/matlab-tomo-2d}{https://github.com/phymhan/matlab-tomo-2d}.}
 In particular,  the middle picture shows radiation sensitivity in projection space, i.e. effective dose caused by a unit of radiation sent into each projection bin. One can see a bright curve strip of high sensitivity. As we will see later, our algorithm will reduce amount of radiation sent into this strip.\par
Building the field design vector $q$ in the model presented in the previous section requires, along with the matrices $A$ and $B$ readily given by the geometry of bins and the ROI, the knowledge of vectors $c$ and $\rho$ with entries indexed by bins.
The entries in $c$  are proportional to our guess on effective dose from a single photon sent into the corresponding bin, and entries in $\rho$ are {estimations of}  the quantities $\exp\{-a_i^Tx_*\}$. Both these vectors depend on the actual image $x_*$. ``In real life'' they {may be either considered available a priori or} should be obtained  from a pilot recovery of $x_*$ in which the radiation is a small fraction of $R$ split, say, equally between $m$ bins.
In our ``proof of concept'' experiment where the goal is to understand potential of radiation design we use ``ideal'' values of $\rho$ and $c$; in particular, we set $\rho_i=\rho^*_i=\exp\{-a_i^Tx_*\}$ . To specify $c$ we act as follows: first, we utilize $x_*$ and pixel sensitivities image $s$ depicted in Figure \ref{fig_phantom} to compute the effective dose $\overline{c}_i$ caused by a single photon sent through the bin $i$. Then, to get $c$, we scale $\overline{c}$ to ensure that $c^T\overline{q}=1$, where $\overline{q}$ is the { uniform (``baseline'') design} such that the radiation in every bin is $R/m$.
\subsection{Computations} Design optimization was carried out using alternating minimization \eqref{G_update}--\eqref{q_update}  described in Section \ref{alternate};
5-6 iterations of the process turned out to be sufficient to get high accuracy solution. An alternative, somehow more time consuming  {solution} is {provided by} the first order algorithm---Mirror Descent with simplex setup, see
\cite[Lecture 5]{ben2021lectures}.
Maximum Likelihood estimate was computed utilizing Nesterov's Fast Gradients
\cite{nesterov2018lectures}
 for minimizing smooth convex functions over the nonnegative orthant.

\subsection{Numerical results} 
\begin{figure}[ht!]
\centering
\includegraphics[width=0.99\textwidth]{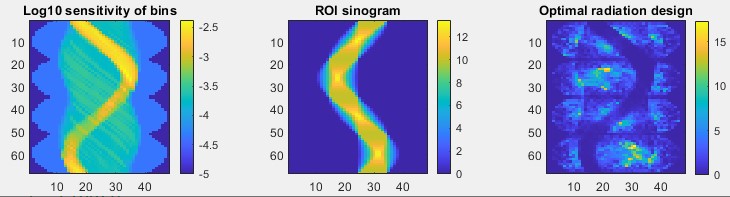}
\caption{Radiation design  for regularization  $\lam$=1e3 (right image.) For comparison, sensitivity and ROI sinograms are presented in the left and the middle plots.
 \label{optimal_radiation_design}}
\end{figure}

\begin{figure}[ht!]
\centering
\begin{tabular}{cc}
\includegraphics[width=0.45\textwidth]{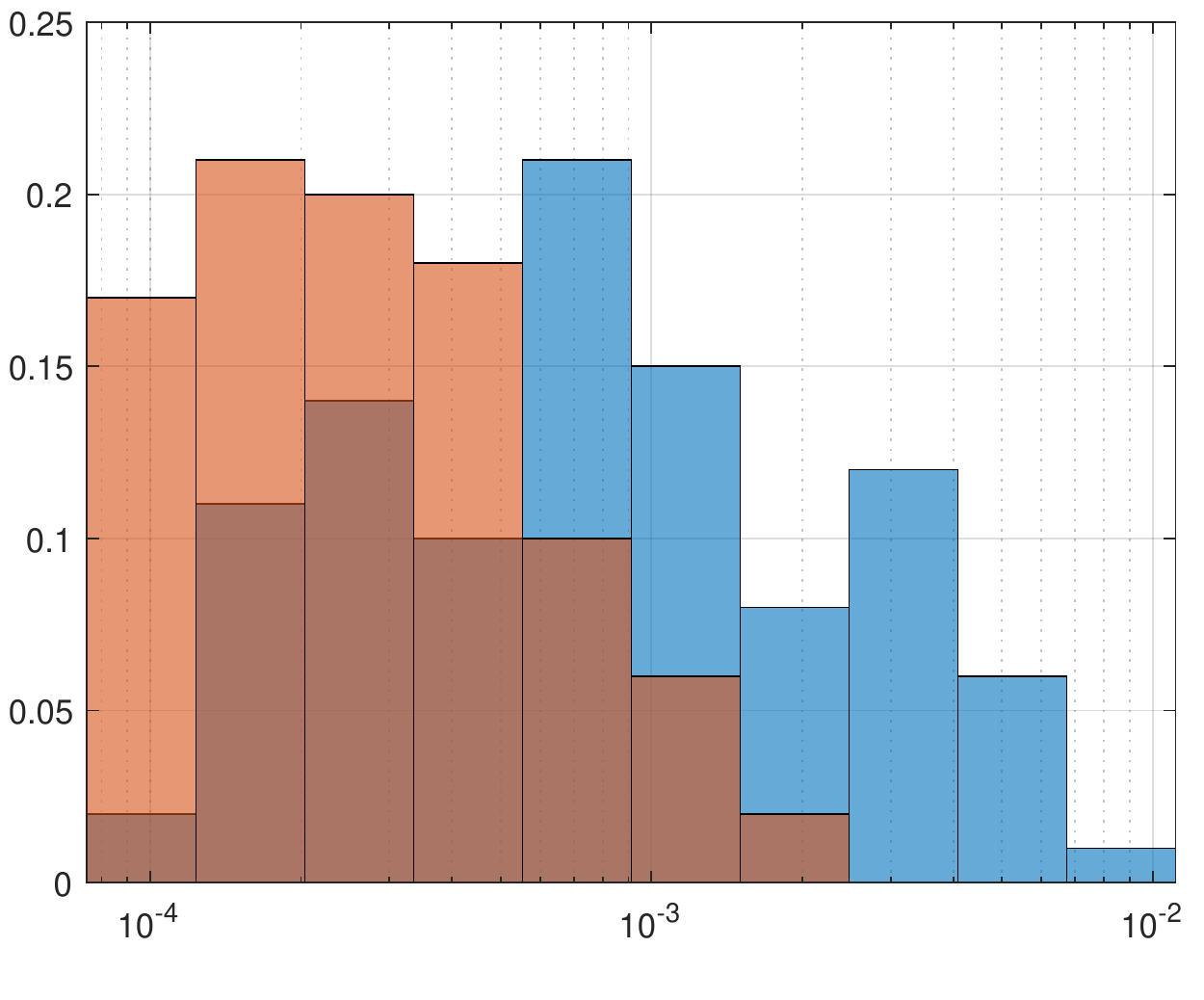}&\includegraphics[width=0.45\textwidth]{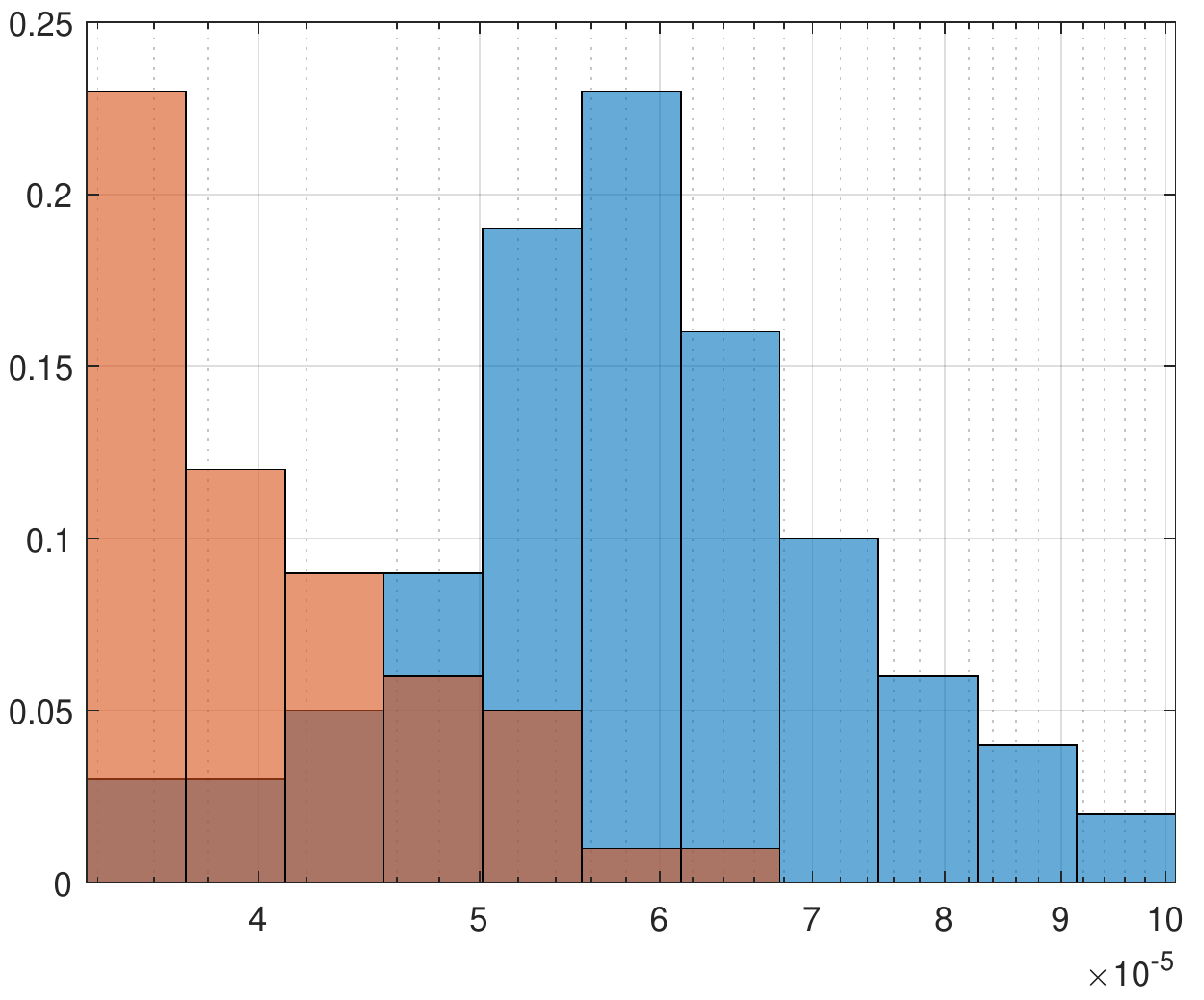}
\end{tabular}
\caption{Distribution of the MSE in experiments with $\lambda=0$ (left plot) and $\lambda=1$e$3$ (right plot): blue histogram---baseline design, red histogram---optimized design.
  \label{roi_MSE}}
\end{figure}

\begin{figure}[ht!]
\centering
\includegraphics[width=0.99\textwidth]{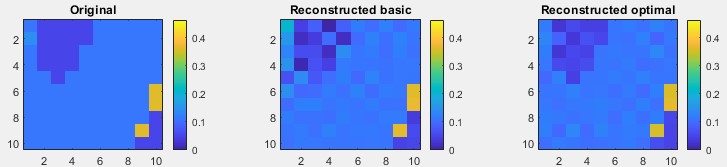}
\caption{Original and reconstructed  ROI ({baseline} and optimized designs), $\lambda=0$. \label{roi_reconstructed}}
\end{figure}

We present results of two experiments  for the phantom image presented in  Figure~\ref{fig_phantom} and two values of the regularization parameter, $\lambda=0$ and $\lambda=1$e$3$. Figure~\ref{optimal_radiation_design} shows the obtained optimized radiation design in the penalized problem. The optimization results in significant reduction of radiation in the sensitive area and its increase in the ROI-related part of the sinogram.
In each experiment we compute $N=100$ recoveries utilizing baseline and optimized designs; in Figure
\ref{roi_MSE} we present the histograms of the mean square error (MSE---squared recovery error per ROI pixel).
Figure \ref{roi_reconstructed} shows typical recoveries of the ROI image in the case of $\lambda=0$. One can observe a clear image quality improvement under the optimized design.

\section{Concluding remarks}\label{sect4}

We present the first attempt to treat CT radiation design within a convex optimization framework. Unlike conventional design, it takes into account
the radiation sensitivity map of the body, therefore minimizing the total effective dose.

\subsection*{Further challenges and opportunities}

\paragraph{3D reconstruction} In the 3D case the number of voxels scales as $n^3$ with single-dimension grid size, while the number of bins grows as $n^4$. This gives much more degrees of freedom for the optimal design than in the 2D case. On the other hand, a faster algorithm is needed to implement radiation design in a reasonable time.

Our approach is especially attractive for use in radiation therapy rooms, where a preliminary radiation sensitivity map of a patient is readily available. It may be used e.g. with kilo-voltage cone-beam CT (kV-CBCT) systems integrated into the gantry of linear accelerators.
 \cite{srinivasan2014applications,wang2008dose}.

One more point to note: usually cone beam CT gantry moves with maximal axial speed so that each body point is exposed to x-ray only once for a given transaxial angle.
It may be better to slow this movement down, allowing multiple axial x-ray angles for each tranaxial angle. This would improve the reconstruction noise-to-dose ratio in CT systems in both cases, with and without control of radiation pattern.

\paragraph{Radiation design under constraints} In this work the design is optimized using an asymptotic maximum likelihood reconstruction model under unique effective dose constraint
(\ref{cq}). One can easily impose extra convex constraints on the radiation vector $q$ (e.g., total radiation dose, maximal dose per pixel, etc). Furthermore, the proposed design optimization framework can be adapted to account for the available a priori information about the body image when it is formulated in the form of convex constraints on the image space (cf., e.g., \cite[Section 6.4.2]{juditsky2020statistical}).

\appendix

\section{Proof of Proposition \ref{propo}}\label{app.prop1}
Given $m\times n$ matrix $P$ of rank $n$ and $\lambda>0$, consider the optimization problem
$$
\min_{H\in\bR^{k\times m}}\left\{\lambda^{-1}\|HP-B\|_F^2+ \|H\|_F^2\right\}
$$
 The objective in this unconstrained minimization problem is a strongly convex quadratic function, so that optimal $H$ is given the unique
  solution to the Fermat equation
$$
\lambda^{-1}P[P^TH^T-B^T]+H^T=0,\
$$
resulting in \[
H=B[\lambda^{-1}P][\lambda^{-1}PP^T+I_m]^{-1}=B[P^TP+\lambda I_n]^{-1}P^T,
\] where the second equality is due to Sherman-Morrison formula. Direct computation shows that for this $H$ one has \[\lambda^{-1}\|HP-B\|_F^2+\|H\|_F^2=\Tr(B[P^TP+\lambda I_n]^{-1}B^T).
\]
Setting  $P=W^{1/2}A$ and substituting the optimization variable $H$ with $GW^{-1/2}$, we arrive at (\ref{poslambda.a}).  Passing in the latter relation to limit as $\lambda\to+0$ and taking into account that $A$ is of rank $n$, we arrive at (\ref{poslambda.b}). \qed

\end{document}